\documentclass{article}
\usepackage{authblk} 
\usepackage{graphicx}
\usepackage[utf8]{inputenc}
\usepackage[T1]{fontenc}
\usepackage{bm}
\usepackage[a4paper, total={7in, 8in}]{geometry}
\usepackage{multicol}
\usepackage{float}
\usepackage{blindtext}
\usepackage{siunitx}
\usepackage[superscript, biblabel, nomove]{cite}
\usepackage{xcolor}
\usepackage{amsmath}
\usepackage{amsfonts} 
\usepackage{mathtools}
\usepackage{nccmath}
\usepackage{verbatim}
\usepackage{longtable}
\usepackage{booktabs}
\usepackage[normalem]{ulem}
\usepackage{textgreek}
\usepackage{textcomp}

\newcommand{\wn}{\textrm{cm}$^{-1}$}

\title{In-plane vector-field imaging of propagating surface phonon polaritons}

\author[1,2,$\dag$]{Richarda Niemann}
\author[1]{Dorothée S. Mader}
\author[3,1]{Gonzalo Alvarez-Pérez}
\author[4,*]{Ana I. F. Tresguerres-Mata}
\author[4]{Aitana Tarazaga Martín-Luengo}
\author[5]{Stefan Partel}
\author[2]{Joshua D. Caldwell}
\author[6,1]{Niclas S. Mueller}
\author[1]{Martin Wolf}
\author[4]{Javier Martín-Sánchez}
\author[4]{Pablo Alonso-González}
\author[1,$\dag$]{Alexander Paarmann}

\affil[1]{Fritz Haber Institute of the Max Planck Society, Berlin, Germany}
\affil[2]{Vanderbilt University, Nashville, TN, USA}
\affil[3]{Italian Institute of Technology, Lecce, Italy} %
\affil[4]{University of Oviedo, Oviedo, Spain}
\affil[5]{Vorarlberg University of Applied Sciences, Research Center of Microtechnology, Dornbirn, Austria}
\affil[6]{Freie Universität Berlin, Department of Physics, Berlin, Germany}
\affil[*]{Present address: Institut Langevin, ESPCI Paris, PSL University, CNRS, Paris, France}
\affil[$\dag$]{Corresponding authors: richarda.niemann@vanderbilt.edu, alexander.paarmann@fhi-berlin.mpg.de}

\begin{document}

\flushbottom
\maketitle

\begin{abstract}
Polariton interferometry through optical near-field microscopy has become a powerful tool in nanophotonics, enabling direct spatial access to the propagation characteristics of strongly confined, evanescent polariton modes. Scattering-type near-field optical microscopy has matured as the prime tool for such studies, yet mostly the out-of-plane components of the optical near fields are probed, owing to the elongated geometry of the nanotip. Here, we demonstrate a complementary far-field nonlinear microscopy approach which allows to selectively probe in-plane polariton field components. Accessing the full vector field is interesting when studying complex mode patterns such as hyperbolic polaritons or skyrmions, where the in-plane field components are typically only inferred from the out-of-plane component but not measured directly . To this end, we use nonlinear infrared-visible wide-field sum-frequency generation microscopy, where the short visible wavelength of the nonlinear signal 
provides the high spatial resolution to access evanescent modes in the infrared. The symmetry selection rules of the nonlinear process further enable polarization-selective imaging of both in-plane polariton field components through spatial interferometry. The concept is demonstrated experimentally using surface phonon polaritons at the AlN-air interface launched by a gold antenna. A simple, semi-analytical model reproduces the peculiar propagation patterns. Hyperspectral imaging with a tunable narrowband laser  further gives access to the polariton dispersion.. The wide-field methodology holds high promise for in-depth and high-throughput studies of infrared nanophotonic structures.
\end{abstract}

\begin{multicols}{2}
    
\section{Introduction} 
\label{intro}

Infrared (IR) nanophotonics has seen a dramatic evolution over the last decade due to the rediscovery of phonon polaritons\cite{caldwell2015low} - hybrid photon-phonon quasiparticles -  in polar crystals. The fundamental properties of surface phonon polaritons (SPhPs) 
 had already been discussed in  literature.\cite{barker1972direct,marschall1972dispersion,falge1973dispersion} Recently, however, phonon polariton research was largely driven by the discovery of natural hyperbolicity\cite{dai2014tunable,caldwell2014sub}  - the most extreme case of optical anisotropy with opposite signs of the permittivity along different crystal axes.\cite{galiffi2024extreme,zhou2025fundamental,he2021phonon,ma2018plane,ma2021ghost,passler2022hyperbolic,ni2023observation,diaz2025visualization} Yet, also SPhPs supported at interfaces and in nanostructures of isotropic or nearly-isotropic polar crystals have been studied extensively,\cite{caldwell2013low,barnett2022investigation,carini2025surface} due to their potential to control thermal transport,\cite{pan2023remarkable} thermal emission,\cite{lu2021engineering} or molecular sensing.\cite{folland2020vibrational} Many different approaches to control the nanoscale polariton propagation have been demonstrated including photo-injection,\cite{dunkelberger2018active} electrical gating,\cite{hu2023gate} twist-optics,\cite{hu2020topological,duan2020twisted,alvarez2024unidirectional} intercalation,\cite{taboada2020broad} isotopic substitution,\cite{giles2018ultralow,chen2023van,carini2026spectral} and symmetry breaking.\cite{hu2023source,matson2023controlling}

\begin{figure*}[ht!] 
\centering\includegraphics[width=0.5\textwidth]{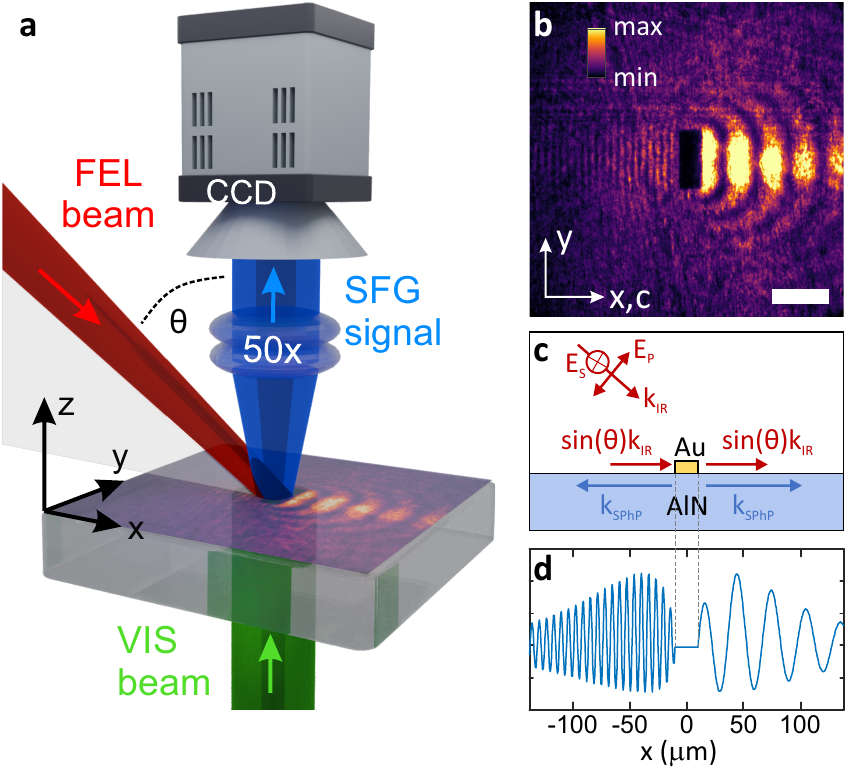}
    \caption{\textbf{Sum-frequency polariton imaging}. (a) Schematic of the SFG microscopy setup. On the front-side, an obliquely incident IR-FEL beam is overlapped with a backside-illuminating VIS beam to generate SFG from the AlN single crystal surface. The SFG signal is collected using a 50x objective and imaged onto a CCD camera. (b) SFG image at $\omega_{\text{IR}}$=792~\wn\,showing fringes from SPhPs propagating along the AlN-air interface, launched by a rectangular gold patch. The IR beam is p-polarized and incident from the left side of the image. The optical $c$-axis of the $m$-cut AlN was parallel to $x$ in this case. The scale bar corresponds to 50 $\mu$m. (c) Schematic of the interference mechanism that leads to polariton fringes in the image, resulting from superposition of the obliquely incident IR beam (red) and the polaritons launched by edges of the gold patch (blue).
    (d) Simulation of co-propagation (right side) and counter-propagation (left side) of these two waves that result in a slow and fast SFG intensity modulation, respectively.} 
    \label{fig:fig1}
\end{figure*}

Notably, the large majority of recent studies of strongly confined phonon polaritons employ scattering-type scanning near-field optical microscopy (s-SNOM)\cite{hillenbrand2025visible} as the major experimental tool. Indeed, s-SNOM provides unprecedented spatial resolution of up to 10~nm\cite{hillenbrand2025visible} providing access to the most confined polariton modes. However, due to the nanotip-sample geometry in s-SNOM, these experiments typically largely probe the normal-to-surface component of the optical near fields, with much reduced sensitivity to in-plane field components.\cite{yao2021probing,neuman2015mapping} It is interesting to note that in visible plasmonics, several microscopy approaches based on nonlinear optics rather than scanning probe techniques have been demonstrated, where authors make use of the polarizations of the ultrashort laser pulses in order to map the plasmonic in-plane vector-fields in the far-field.\cite{davis2020ultrafast,frischwasser2021real} It is intriguing to transfer these approaches to the IR and phonon polaritons. To that end, we have recently introduced sub-diffractional imaging of SPhPs via sum-frequency generation (SFG) microscopy,\cite{niemann2022long,niemann2024spectroscopic} providing sufficient spatial resolution to image confined modes and polariton contrast through the intrinsic sensitivity to the polaritonic local IR field enhancements. In these experiments, we employed c-cut 4H-SiC as a polar crystal supporting propagating and localized SPhPs. Due to the SFG selection rules in  this specific configuration, however, the signals in these experiments exclusively emerge from the out-of-plane polaritonic field components,\cite{niemann2022long,niemann2024spectroscopic} just like in s-SNOM, with no sensitivity to in-plane fields.

In this work, we experimentally demonstrate in-plane vector-field imaging of propagating SPhPs at the AlN-air interface. We employ wide-field, infrared-visible (IR-VIS) SFG  microscopy\cite{niemann2022long,niemann2024spectroscopic,mader2024sum,mueller2026full,khan2023compact}, see Fig.~\ref{fig:fig1}a. The instrument uses an objective and CCD camera to image the up-converted nonlinear SFG photons that leave the sample after frequency-mixing of the IR and VIS fields. 
We placed gold patches which act as antennas on an m-cut (010) AlN crystal surface, to launch SPhPs from the antenna edges. With this arrangement, we observe interference of the propagating polaritons with the illuminating IR beam. Specifically for m-cut AlN, the selection rules of the nonlinear process allow to selectively probe the in-plane IR field components parallel (x) and perpendicular (y) to the incidence plane (cf. Fig.~\ref{fig:fig1}a for the definition of the coordinate system) such that the full polariton in-plane vector-field can be accessed. A simple analytical model of polariton point sources placed along the gold antenna edges accounting for phase retardation across the field-of-view
, and the interference of the resulting total polariton field with the incident IR field reproduces the unique propagation features observed in the experiment. Furthermore, the rapid data acquisition enables spectroscopic imaging to provide the full polariton dispersion.

\section{Polariton vector-field imaging by sum-frequency generation} 
\label{vector-field}

The experimental setup of the SFG microscope depicted in Fig.~\ref{fig:fig1}a has been characterized in detail in ref. \citenum{niemann2022long}, and is described in the Methods. In short, a tunable, narrowband IR free-electron laser (IR-FEL)\cite{schollkopf2015new,schoellkopf2026twocolordualoscillatorinfraredfreeelectron} under oblique incidence in the $x-z$  plane, is spatially overlapped with a VIS laser normally incident from the back side of the sample, to generate a sum-frequency signal in the visible spectrum, $\lambda_\text{SFG} \approx 500~$nm. We use a 50x objective to image the SFG signal onto a CCD camera, yielding a spatial resolution of $\approx$1.4~\textmu m. We used AlN, a uniaxial polar crystal with two infrared-active phonon modes of
different symmetry \cite{moore2005infrared}: the $E_1$ mode, with atomic displacements perpendicular to the optical $c$-axis, and the $A_1$ mode, with displacements along the $c$-axis. Each mode gives rise to a distinct Reststrahlenband in the corresponding dielectric component, $\varepsilon_\perp$ and $\varepsilon_\parallel$ respectively, resulting in a strongly anisotropic optical response in the mid-infrared. For the $m$-cut crystal used here, the $c$-axis lies in the surface plane, parallel to $x$. A detailed discussion of the optical response of AlN and the resulting SPhP dispersion is given in Supplementary Section S1. We placed $20 \times 50$ \textmu m$^2$ gold antennas onto an m-cut AlN single crystal with the short antenna axis parallel to the AlN in-plane optical $c$-axis, see Methods for sample fabrication details.

A typical SFG image of propagating SPhPs within the elliptic Reststrahlenband of AlN\cite{moore2005infrared} is shown in Fig.~\ref{fig:fig1}b for $\omega_{\text{IR}}=792$~\wn. A pronounced fringe pattern is observed around the rectangular gold antenna in the middle of the image, with long period fringes on the right side and short period fringes on the left side of the antenna.  The fringes emerge from two interfering waves as schematically shown in Fig.~\ref{fig:fig1}c: (red) the incident IR field propagating  along $x$ with the projected in-plane momentum $\sin(\theta)k_{\text{IR}}$, and (blue) the surface polariton propagating backwards with $-k_{\text{SPhP}}$ on the left side of the antenna and forward with $+k_{\text{SPhP}}$ on the right side of the antenna. The superposition of both (time-varying) electric fields results in a time-independent intensity modulation, i.e., a standing wave as shown in Fig.~\ref{fig:fig1}d, with short-wavelength and long-wavelength modulations in the counter- and co-propagating directions, respectively. This can be expressed by looking at the field intensity $|E|^2$ that will show a spatial modulation arising from the interference between both the IR and SPhP waves:
\begin{eqnarray} \label{eq:interference}
   |E|^2 &=& |E_{\text{IR}}(x) + E_{\text{SPhP}}(x)|^2 \\ 
    &=& |E_{\text{IR}(x)}|^2 + |E_{\text{SPhP}}(x)|^2 \nonumber \\
    & & + 2|E_{\text{IR}}(x)||E_{\text{SPhP}}(x)| \sin(\Delta k_\pm \cdot x + \phi_0)
   \nonumber
\end{eqnarray} 
where $\Delta k_\pm = \sin\theta k_{\text{IR}} \mp k_{\text{SPhP}}$ describes the momentum difference between co-propagating (+) and counter-propagating (-) waves, $|E_{\text{IR}}(x)|$ and $|E_{\text{SPhP}}(x)|$ their envelopes, and $\phi_0$ a phase offset between them. Note that, following Eq.~\eqref{eq:interference}, counter-propagating waves result in fringes corresponding to the sum of both momenta, while co-propagating waves lead to fringes corresponding to their momentum difference. Thus, the fringe spacing for co-propagation can easily be larger than the polariton wavelength and even the free-space wavelength, as is the case on the right-hand side of Fig.~\ref{fig:fig1}b. A similar observation was reported using s-SNOM for low-confinement, near-light-line polaritons.\cite{barnett2022investigation}

To comprehend why the SFG image indeed emerges from a specific in-plane polariton field  component, it is useful to understand the tensorial nature of the nonlinear-optical sum-frequency process. The polarized sum-frequency response of m-cut AlN has been investigated in detail before.\cite{mader2024sum} In short, in the given geometry, each of the four polarization configurations (SP, PS, PP, and SS, s- or p-polarization each for VIS and IR beams, respectively) only yields a single contribution with a non-vanishing component of the second-order nonlinear susceptibility tensor $\chi^{(2)}$, all of which only contain in-plane electric field components of the IR beam. For the specific case of  $x || c$, the polarized SFG responses read:\cite{mader2024sum} 
\begin{eqnarray} \nonumber
    I^{\text{SFG}}_{\text{SP}} \propto |\chi^{(2)}_{\text{aac}} E_{\text{VIS,y}} E_{\text{IR,x}}|^2, \\ \nonumber
    I^{\text{SFG}}_{\text{PS}} \propto |\chi^{(2)}_{\text{aca}} E_{\text{VIS,x}} E_{\text{IR,y}}|^2, \\ \nonumber
    I^{\text{SFG}}_{\text{PP}} \propto |\chi^{(2)}_{\text{ccc}} E_{\text{VIS,x}} E_{\text{IR,x}}|^2, \\ \label{eq:SFGpol}
    I^{\text{SFG}}_{\text{SS}} \propto |\chi^{(2)}_{\text{caa}} E_{\text{VIS,y}} E_{\text{IR,y}}|^2.
\end{eqnarray}
Effective $\chi^{2}$ values emerge for non-trivial azimuthal rotations, while maintaining the same simple functional form, see SI for details. Eqs.~\eqref{eq:SFGpol} provide strict selection rules for which IR field components contribute to the SFG signal in a given polarization configuration of the exciting laser fields for a homogeneous AlN-air interface.\cite{mader2024sum} Scattering of the incident IR field at the gold antenna, however, is expected to also result in partial polarization conversion, for instance by launching a polariton. Such polarization conversion will lead to an - at least partial - lifting of the selection rules, potentially preventing selective polariton vector-field imaging if both in-plane polariton field components contribute to an SFG image. However, the current approach explicitly relies on spatial interference of a mixed-polarization-state polariton field with the fully polarized incident field, Eq.~\eqref{eq:interference}. Thus, the polariton fringes observed in the SFG image such as in Fig.~\ref{fig:fig1}b exclusively emerge from one specific in-plane polariton field component, only. In other words, polariton interferometry in SFG imaging essentially projects the polariton fields onto the well-defined, polarized incident field direction, and thus  enables polariton vector-field analysis. 

\begin{figure*}[htb!] 
\centering\includegraphics[width=1.0\textwidth]{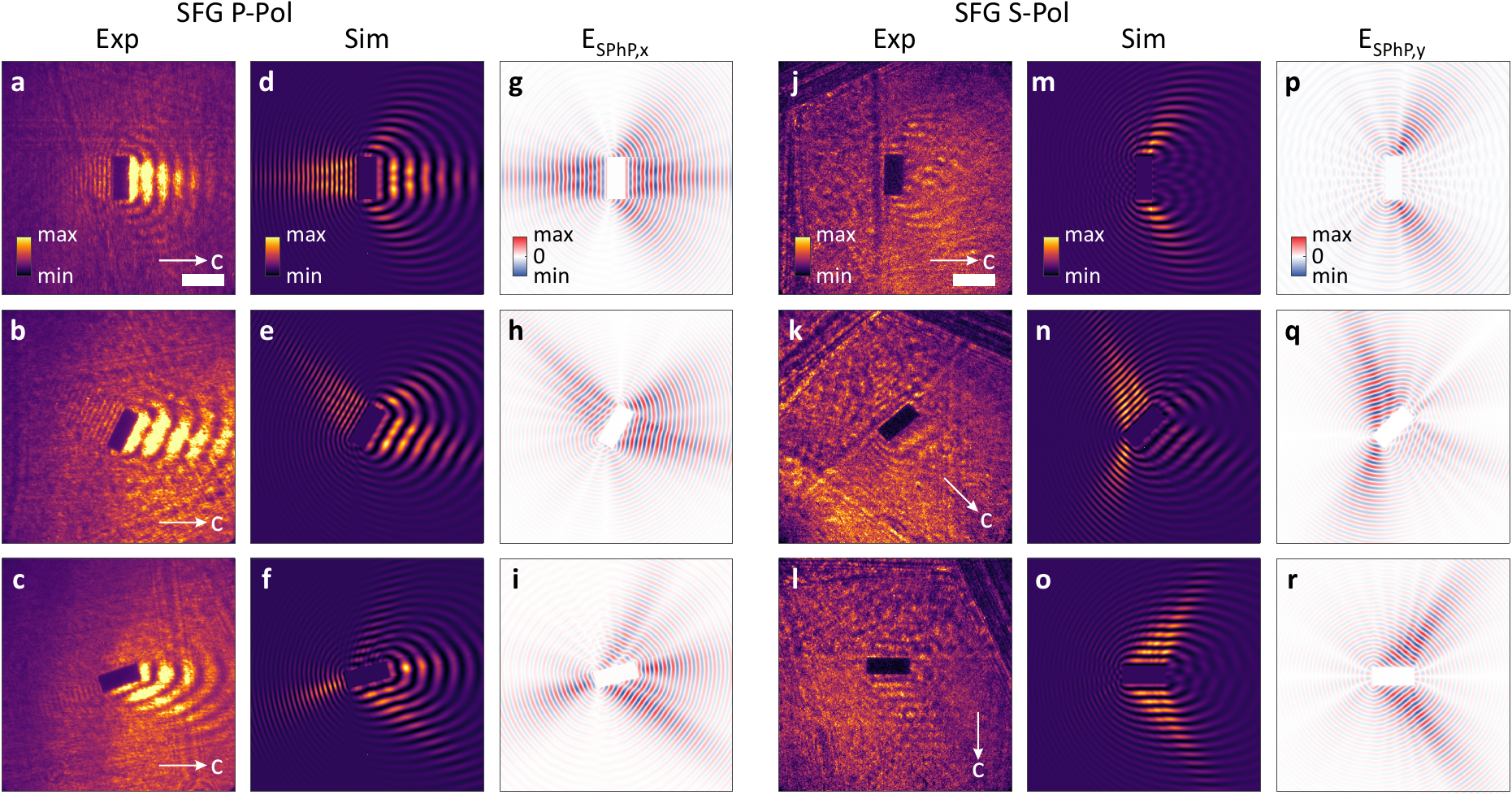}
    \caption{\textbf{Vector-field imaging of SPhPs.} (a-c,j-l) Experimental wide-field SFG images for different gold patch orientations under p-pol (a-c) and s-pol (j-l) IR illumination, resulting in the images arising from the $x$ and $y$ field components of the polaritons, respectively. (d-f,m-o) Simulated SFG images for the corresponding experimental configurations. (g-i,p-r) Associated simulated polariton $x$ (g-i) and $y$ (p-r) field components. All measurements and simulations shown are taken at $\omega_{IR}=820$~cm$^{-1}$. The scale bar shown in (a,j) applies to all subplots and corresponds to 50~\textmu m, and the white arrow marks the direction of the optical $c$-axis of AlN for each image. Please note that the significantly reduced scattering efficiency at the gold patch edges under s-pol excitation leads to reduced signal-to-noise in these images.}
    \label{fig:fig2}
\end{figure*}

To fully explore the concept of vector-field SPhP imaging, we carried out hyperspectral SFG microscopy for SP and PS polarization configurations, and for a series of antenna orientations, see Supplementary Videos. In Fig.~\ref{fig:fig2}, we show a subset of these data for a single IR excitation frequency ($\omega_{IR}$ = 820~\wn), for SP (left side) and PS (right side) polarization configurations, and selected sample azimuths. We first focus on the case of a p-polarized IR beam, Fig.~\ref{fig:fig2}a-c. Fig.~\ref{fig:fig2}a reproduces the features seen in Fig.~\ref{fig:fig1}b, just with slightly shorter fringe spacings overall due to the shorter wavelengths of excitation as well as the AlN SPhP. Markedly, apart from propagation along $x$, there is also a weaker feature of polariton propagation along $y$, mostly into the forward direction and disappearing directly above the antenna. To understand this qualitatively, we can assume the antenna edges to be represented by a series of SPhP point sources emitting circular surface waves with radial electric fields, which means for propagation along $x$ we expect polariton fields in the $x-z$ plane, while for propagation along $y$ fields in the $y-z$ plane.\cite{passler2017generalized} Therefore, the disappearance of fringes along the $y$-direction directly above the antenna does not necessarily mean that there is no more polariton field, but instead that the in-plane polariton field is now polarized perpendicular to the incident field $E^{\text{P}}_{\text{IR}}=(E_{\text{IR,x}},0,E_{\text{IR,z}})$ and, thus, spatial interferometry is suppressed. 

Additionally, the finite size and rectangular shape of the antenna will also modify the polariton field distribution, due to interference between all the scattering sources along the antenna edges. This is apparent when rotating the antenna by 30° and 75°, see Fig.~\ref{fig:fig2}b and c, respectively. Interestingly, a peculiar effect is observed in these images, most pronounced in Fig.~\ref{fig:fig2}c. The fringes emerge parallel to the antenna edge, yet the overall energy flow seems to be significantly tilted, suggesting a misalignment between momentum and Poynting vector for these SPhPs. While such a behavior is common for in-plane hyperbolic polaritons,\cite{voronin2025misalignment,voronin2024fundamentals,zhou2025fundamental} at the excitation frequency used here, $\omega_\text{IR} = 820$~cm$^{-1}$, the anisotropy is small ($\Delta\epsilon \approx 1.1$, see Supplementary Section 1). The isofrequency curve at this frequency is therefore nearly circular and the misalignment between the SPhPs wavevector and direction of propagation amounts to only a few degrees (see Supplementary Figure 1), which cannot explain the observation. Instead, we find that this phenomenon emerges from the spatial interference, and in particular from the relative phase at which the polaritons are launched at different positions along the gold edge.  In order to explain the image formation, we set up a simple semi-analytical model, the results of which are shown in Fig.~\ref{fig:fig2}d-f and m-o for p- and s- polarized IR light, respectively. In this model, we place point sources of circular SPhP waves closely spaced along the edges of the gold antenna. 

For each of these circular waves, the in-plane polariton field is assumed to be parallel to the propagation direction. To mimic the experimental situation, only the half-space pointing away from the gold patch is included in the circular wave, assuming that the other half-space is efficiently damped to not reemerge on the other side of the antenna, see Suppl. Fig. S2 for an illustration of this procedure. 
Importantly, we further assume that each point source along the antenna edge inherits the phase of the incident IR field, see Methods for a full description of the model. 

For the vertical antenna in Fig.~\ref{fig:fig2}a, this simple model results in polariton $x$ and $y$ field component profiles as shown in Fig.~\ref{fig:fig2}g,p for p- and s-polarized IR light, respectively. Interfering these polariton fields with the corresponding incident IR fields according to Eq.~\eqref{eq:interference} results in simulated polariton SFG maps as shown in Fig.~\ref{fig:fig2}d,m. Overall, our simulations reproduce the experimentally observed mode patterns with remarkable precision, across all the arrangements shown in Fig.~\ref{fig:fig2}. Markedly, the model also reproduces the fringes to emerge parallel to the patch edges but with energy flow into a sloped direction, in agreement with the experiment. Inspecting the polariton fields (cf. Fig.~\ref{fig:fig2}, however, it becomes clear that the polariton itself still shows 'normal' behavior expected for an isotropic system, in particular the polariton phase fronts are perpendicular to the direction of energy flow, see Fig.~\ref{fig:fig2}g-i,p-r. We note that, indeed, the s-polarized IR laser is significantly less efficient in exciting polaritons at the gold edges, leading to lower-quality SFG images for this configuration, Fig.~\ref{fig:fig2}j-l. Nonetheless, the main propagation features as predicted in Fig.~\ref{fig:fig2}m-o can still be well-recognized also in these lower-quality data, confirming that the observed patterns indeed emerge from the interference of the incident field and
that launched by the antenna. Note also, that for the s-pol measurements, the whole sample was rotated which, however, does not significantly affect the SFG images. Overall, the excellent agreement between experiment and simulation in Fig.~\ref{fig:fig2} proves that our experiment is indeed probing the in-plane polariton fields selectively.   

\begin{figure*}[ht!] 
\centering\includegraphics[width=0.6\textwidth]{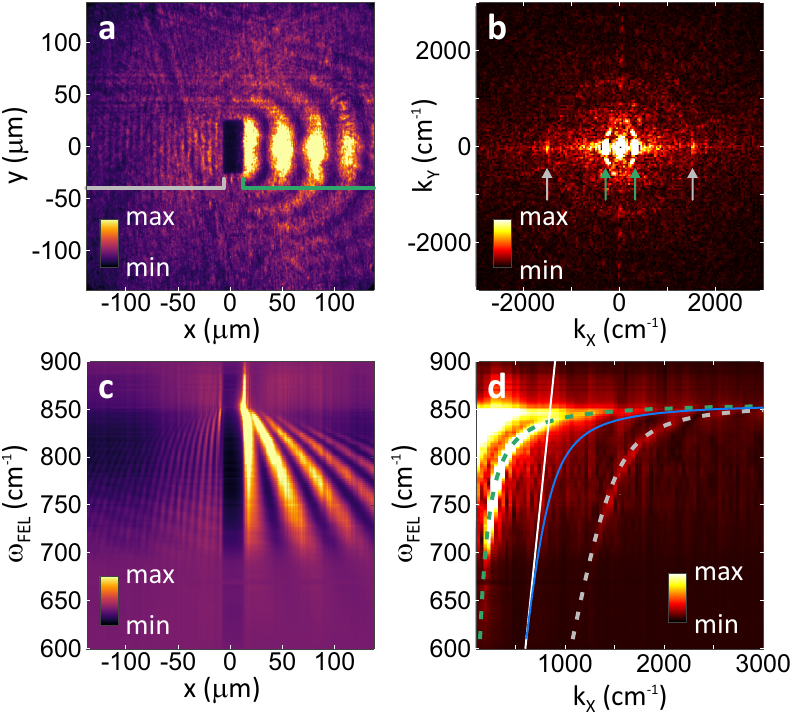}
    \caption{\textbf{Spectroscopic Imaging and Polariton Dispersion}. (a) SFG image at $\omega_{\text{IR}}=780$~\wn,  illustrating the co-propagating (green) and counter-propagating (gray) polariton contributions. (b) 2D-FFT of the image shown in a), where the co- and counter-propagating wave signatures are marked with arrows. (c) Spectral dependence of the polariton propagation extracted by $y$-integration of the SFG image for each IR excitation frequency. (d) 1D-FFT for each frequency of the propagation patterns shown in c). Overlaid are the predicted co- and counter-propagating dispersions (green and gray, respectively), the native polariton dispersion (blue), as well as the light line (white). Note that the reduction of the SFG signal around $\omega_{IR}\approx 670$~\wn arises due to CO$_2$ absorption of the IR beam in ambient air.}
    \label{fig:fig3}
\end{figure*}


In order to quantitatively extract the polariton dispersion, we further analyzed the spectroscopic SFG imaging data for the vertically aligned gold patch, see Fig.~\ref{fig:fig3}a for an example. The co- and counter-propagating polariton fringes are marked in green and gray, respectively. 
In particular, these features can also be recognized as peaks in the two-dimensional fast-Fourier transformed (2D-FFT) image shown in Fig.~\ref{fig:fig3}b. To simplify the analysis, we reduced the dimensionality of the problem by integrating the SFG images along $y$ across the height of the gold antenna (see Methods for details
). 
In Fig.~\ref{fig:fig3}c, the resulting IR frequency-dependent 1D propagation patterns are shown. These data very nicely illustrate the dispersion of the SPhP in real space across the whole AlN Reststrahlenband. By performing 1D-FFT for each IR frequency, the full dispersion of the SPhP is recovered as shown in Fig.~\ref{fig:fig3}d. 

Notably, our simple model to reproduce the SPhP propagation patterns at specific IR frequencies as shown in Fig.~\ref{fig:fig2} did not depend sensitively on the details of the AlN SPhP dispersion, and in particular on its anisotropy. In fact, for uniaxial polar crystals, three different surface modes emerge along high-symmetry directions, depending on the alignment of the optical axis with regard to the surface plane and the polariton propagation direction\cite{falge1973dispersion}:
\begin{eqnarray} 
    \label{eq:oSPhP}
    k_{\text{o,SPhP}} &=& \Re  \left[  \sqrt{\varepsilon_\perp^\text{AlN}/(\varepsilon_\perp^\text{AlN} + 1) }\, \right], \\
    k_{\text{eoI,SPhP}} &=& \Re  \left[  \sqrt{\frac{\varepsilon_\parallel^\text{AlN}\varepsilon_\perp^\text{AlN}-\varepsilon_\perp^\text{AlN}}{\varepsilon_\parallel^\text{AlN}\varepsilon_\perp^\text{AlN} - 1} }\, \right],
    \label{eq:eoISPhP} 
    \\
    k_{\text{eoII,SPhP}} &=& \Re  \left[  \sqrt{\frac{\varepsilon_\parallel^\text{AlN}\varepsilon_\perp^\text{AlN}-\varepsilon_\parallel^\text{AlN}}{\varepsilon_\parallel^\text{AlN}\varepsilon_\perp^\text{AlN} - 1} }\, \right],
    \label{eq:eoIISPhP}
\end{eqnarray}
corresponding to polarization perpendicular to the $c$-axis (Eq.~\ref{eq:oSPhP}, isotropic-equivalent SPhP\cite{caldwell2015low}), $c$-axis in the surface plane ($E_{SPhP}^x \parallel c$, Eq.~\ref{eq:eoISPhP}), and $c$-axis normal to the surface ($E_{SPhP}^z \parallel c$, Eq.~\ref{eq:eoIISPhP})\cite{falge1973dispersion}. In spectral regions of small optical anisotropy, however, as is the case for the data shown in Fig.~\ref{fig:fig2}, these three SPhP dispersions for AlN are nearly identical, and we could well-reproduce the experimental images using the simplest, isotropic dispersion model, Eq.~\ref{eq:oSPhP}.

However, in order to accurately describe the full experimental dispersion in Fig.~\ref{fig:fig3}d, we do need to account for the anisotropy of $m$-AlN, as it spans the entire Reststrahlenband, including frequencies near the high-frequency SPhP cut-off $\omega_s$\cite{caldwell2015low,falge1973dispersion}. In this regime, the isotropic approximation used for the simulations shown in Fig. \ref{fig:fig2} breaks down, and SPhP propagation is well described only when the anisotropic dispersion is taken into account. For the optical $c$-axis of the uniaxial crystal in the surface plane and polaritons propagating along this $c$-axis, that is, the dispersion along the axis of anisotropy, we expect the dispersion for extraordinary polaritons Eq.~\ref{eq:eoISPhP}
shown in blue in Fig.~\ref{fig:fig3}d. When additionally accounting for the co-and counter-propagating interference (Eq.~\ref{eq:interference}) shown in green and gray, respectively, which displaces the dispersion by $\pm\Delta k = \pm\sin(\theta)k_{\textrm{IR}}$, we observe near-perfect agreement with the experiment, confirming that our SFG imaging approach gives quantitative access to the in-plane polariton fields and their dispersion. Notably, using wide-field microscopy we here attain these data at a rapid acquisition speed, orders-of-magnitude faster than using point-scanning approaches like s-SNOM, such that a full hyperspectral scan is typically acquired in 30~mins,\emph{i.e.} similar to a typical single-frequency image acquisition time in s-SNOM.


\section{Discussion and Outlook}
\label{discussion}

In this work, we experimentally demonstrate sub-diffractional vector-field imaging by means of SFG spectro-microscopy. The  selection rules in nonlinear optics enabled us to selectively image the in-plane $x$ and $y$ polariton field components by employing p- and s-polarized IR excitation, respectively. This is in contrast to vector-field imaging of plasmon polaritons at metallic surfaces using photo-electron emission microscopy (PEEM). This approach works similarly to out work for s-polarized excitation to measure the $y$-field component, while p-polarized excitation will result in interferences with in- and out-of-plane polariton field components simultaneously, complicating the extraction of the individual field components.\cite{dreher2024spatiotemporal}
It would potentially be very interesting to apply our vector-field imaging approach to more complex excitations such as recently reported SPhP skyrmions\cite{mangold2026phonon} or polaritonic metasurfaces.\cite{niemann2024spectroscopic,gudalla2021topological} Furthermore, it is appealing 
to be able to decouple the excitation of the nanophotonic mode from the interferometric probing of a selected field component. One option to achieve this could be, for instance, to excite polaritons with one IR polarization but probe with crossed polarization in an IR pump-SFG probe scheme. Using short pulses, this would even enable to study the dynamics, similar to time-resolved PEEM.\cite{dreher2024spatiotemporal} Markedly, the SFG emerges only during temporal overlap of the VIS pulses with the IR pump, such that a time-delayed SFG probe could be very efficient in selectively probing a desired component. Furthermore, even two-color IR experiments have just become feasible with the newly implemented dual-oscillator two-color IR-FEL,\cite{schoellkopf2026twocolordualoscillatorinfraredfreeelectron} that would potentially even allow to explore nonlinear polariton-polariton interactions\cite{hellman2025microscopic} through IR pump-vector-field SFG imaging probe experiments. 

The work presented here exhibits two conceptual limitations: (i) the finite spatial resolution of SFG microscopy\cite{niemann2022long} in comparison to other near-field imaging techniques such as s-SNOM\cite{hillenbrand2025visible} and (ii) symmetry, in particular broken inversion symmetry of the polar crystal that is required for efficient SFG.\cite{liu2008sum} Surely, the spatial resolution of the SFG microscope can still be optimized in principle, but is ultimately limited by the diffraction limit at the SFG wavelength.\cite{niemann2022long} However, the absolute spatial resolution is (nearly) independent of the resonant IR wavelength, such that pushing further towards the THz spectral range would enable THz wide-field microscopy at frequencies even 3~THz with a deeply sub-diffractional resolution beyond $\lambda_{\text{IR}}/100$. THz SFG microscopy could be implemented straight-forwardly at THz FEL facilities like the newly implemented far-IR FEL\cite{schoellkopf2026twocolordualoscillatorinfraredfreeelectron} at the Fritz Haber Institute, FELIX\cite{knippels1998two} or FELBE.\cite{helm2023elbe}, and would provide a important complementary microscopy technique in this technologically relevant spectral range. 
Secondly, in order to transfer the vector-field SFG imaging concept to inversion symmetric materials systems, we recently imaged polaritons through their evanescent fields leaking into a inversion-broken substrate acting as a nonlinear transducer for polariton SFG microscopy.\cite{niemann2026remote} Finally, while the current work employs an IR-FEL, we anticipate that rapidly evolving IR laser technology will boost widespread lab-based SFG microscopy.\cite{mueller2026full,fellows2024spiral} 

\section{Conclusion}
\label{conclusion}
In conclusion, we introduced and experimentally demonstrated in-plane vector-field imaging of SPhPs by means of wide-field SFG microscopy, exemplified at the AlN-air interface. We make use of the symmetry of the nonlinear susceptibility tensor of AlN to selectively probe either the $x$ or the $y$ in-plane components of the polariton fields. Spatially interfering the polariton field with the direct wide-field IR excitation beam leads to standing-wave polariton fringes. A semi-analytical model allows to reproduce the intricate fringe patterns with excellent precision. Furthermore, by employing full spectroscopic SFG imaging across the whole AlN Reststrahlenband, we measure the SPhP dispersion which agrees perfectly with the analytical prediction. Thus, nonlinear SFG microscopy is a versatile tool to selectively probe specific components of the sample response with sub-diffractional imaging resolution and wide-field methodology, holding high promise for a wide range of nanophotonics and materials science applications.

\section*{Methods}

\subsection*{SFG spectro-microscopy}

The SFG microscope, including the IR-FEL and synchronized VIS laser system, as well as the measurement procedure have been described in detail before.\cite{niemann2022long,niemann2024spectroscopic,mader2024sum} The IR-FEL with macro (micro) pulse repetition rate 10\,Hz (55.5\,MHz) was focused mildly (spot size 300~\textmu m) onto the sample surface under an oblique incidence angle of $\theta=50^\circ$. The FEL frequency $\omega_\text{IR}$ was tuned between 650~\wn~and 950~\wn.
The visible laser was either a frequency-doubled high power laser ($\lambda_\text{VIS} = 532$~nm, Monfort M-PICO-LAB Nd:VAN seed laser, 1064 nm, 55.5 MHz, Agilite 569-10 Nd:YAG amplifier, 10 Hz macropulse repition rate, pulse duration $>$10~ps, micropulse energy in the microscope of a few \textmu J) for the p-polarized measurements, or a frequency-doubled amplified fiber laser ($\lambda_{\text{VIS}}=520$~nm, Orange High Power 10, Menlo, 1040 nm, repetition rate 55.5~MHz, pulse energy 10~nJ). In both cases, the VIS beam was mildly focussed to a spot size of 500~\textmu m.

Both lasers (IR and VIS) are synchronized,\cite{kiessling2018femtosecond} and temporally overlapped to generate a sum-frequency signal at $1/\lambda_{\text{SFG}} = 1/\lambda_{\text{VIS}} + 1/\lambda_{\text{IR}}$ with $\lambda_\text{SFG} \approx 500~$nm. After spectral filtering, the SFG signal is imaged using a 50x long working distance objective (Mitutoyo M Plan Apo 50×) onto a gated CCD camera (Teledyne PI-Max-4), yielding a spatial resolution of $\approx1.4$~\textmu m.\cite{niemann2022long} SFG images are acquired by averaging 100 FEL macropulses  at each FEL wavelength, corresponding to 10~s integration time on the camera per image, and typical acquisition times of 30~mins for a full hyperspectral map.   

\subsection*{Sample preparation}

We use commercial m-cut AlN single crystals (Nitride Crystals, Inc.) with a series of $20 \times 50$ \textmu m$^2$ gold antennas that were fabricated with different rotations of the antenna long axis with regard to the AlN in-plane optical $c$-axis. We used m-cut AlN as a polar crystal supporting SPhPs within the Reststrahlband, \emph{i.e.}, between the transversal optical (TO) and longitudinal optical (LO) phonon frequencies. For AlN, $\omega_{TO,\parallel}=608.5$~\wn\,and $\omega_{LO,\parallel}=888.9$~\wn (along the $c$-axis), and $\omega_{TO,\perp}=667.2$ ~\wn and $\omega_{LO,\perp}=909.6$~\wn (perpendicular to the $c$-axis).\cite{moore2005infrared}

Au antennas with lateral dimensions of 50 × 20~$\mu$m were fabricated on AlN by direct-write optical lithography and lift-off. The sample was mounted on a 100 mm Si carrier, coated with a positive photoresist, patterned by laser lithography, developed after post-exposure baking and cleaned by a short O$_2$ plasma descum. A 5~nm Cr adhesion layer and 40~nm Au were then deposited by high-vacuum evaporation, followed by lift-off in acetone.
 
\subsection*{Simulations}

We simulated the SFG images shown in Fig.~2 by placing many SPhP point sources at positions $(x_0,y_0)$ all along the edge of the gold antenna, \emph{i.e.}, without explicitly including the gold antenna but only its geometry. We use a point source spacing of 500~nm along the edge for which  the total pattern already converged. A larger spacing may lead to additional features emerging. Each point source was assumed to be a spherical wave with a radial in-plane field: 
\begin{eqnarray} \nonumber
    E_\text{x}^{(x_0,y_0)}(x,y) &=& \frac{x-x_0}{r}\exp{\left( i k_{\text{SPhP}}r + i\phi_0\right)} \\ \nonumber
    E_\text{y}^{(x_0,y_0)}(x,y) &=& \frac{y-y_0}{r}\exp{\left(i k_{\text{SPhP}}r + i\phi_0\right)}, \\ \nonumber
    r &=& \sqrt{(x-x_0)^2+(y-y_0)^2} \\
    \phi_0 &=& \sin{\theta k_{\text{IR}}x_0} , 
    \label{meth:point}
\end{eqnarray}
where $k_{\text{SPhP}}(\omega)$ was defined in Eq.~\ref{eq:oSPhP}, and $k_{\text{IR}} = 2\pi/\lambda_{\text{IR}}$ is the free-space momentum of the IR beam. In Eq.~\ref{meth:point}, we explicitly assumed the IR beam incident in the $x-z$ plane such that the phase $\phi_0$ only varies with $x$ and is constant along $y$. Notably, the initial phase $\phi_0$ for each spherical wave accounts for the phase of the obliquely incident IR beam being transferred onto the polariton at the given position at the edge of the gold antenna. 

Before superposition of all point sources to form the total polariton field as shown in Fig.~\ref{fig:fig2}g-i,p-r, we remove half the spherical wave that would be propagating towards the antenna by employing a respective spatial filter function $F(x_0,y_0)$, i.e., we only consider the half space pointing away from the antenna. We note that this simplified procedure produces small artifacts from not treating any wave bending around the corners of the rectangular antenna. To reduce this effect, we include corner rounding with a radius of ~1.5~\textmu m. The total polariton field then is written as:
\begin{eqnarray}
    E_{\text{SPhP,a}}(x,y) &=& \sum_{(x_0,y_0)} E^{(x_0,y_0)}_\text{a}(x,y) F(x_0,y_0), 
\end{eqnarray}
with $a = x,y$. To finally generate the simulated SFG images, we calculate the intensity of the superposition of the polariton field with the incident IR field:
\begin{eqnarray} \nonumber
    \text{SFG}_\text{P} &=& \left |E_{\text{SPhP,x}}(x,y) + E_{\text{IR}} \right |^2 \\ 
    \text{SFG}_\text{S} &=& \left |E_{\text{SPhP,y}}(x,y) + E_{\text{IR}} \right |^2,
    \\ \nonumber
    E_{\text{IR}} & \propto & \exp{\left(i\sin\theta k_{\text{IR}}x\right)}
\end{eqnarray}

We note that we ignore the anisotropy of AlN for these simulations, which is a rather small effect that would influence the fringe spacing along the $x$-direction. Nonetheless, it would be straight-forward to extend the model to also include the anisotropy or even hyperbolicity, by replacing the circular waves by elliptic or hyperbolic waves, respectively. 

\subsection*{Dispersion extraction}

In order to extract the dispersion from the SFG spectro-microscopy data, we used the spectroscopic data set (series of SFG images for different infrared frequency $\omega_{IR}$) where the antenna long axis was aligned along $y$ (cf. Fig.~\ref{fig:fig2}a). For each image, we integrated the SFG signal along $y$ across the height of the antenna, only, resulting in a line plot along $x$ for each FEL frequency. These line plots are shown in Fig.~\ref{fig:fig3}c. Finally, each of these line plots is Fourier-transformed, as shown in Fig.~\ref{fig:fig3}d.

\section*{Acknowledgements}
R.N. was supported by the United States Army Research Office (ARO) under MURI grant (W911NF2420195). D.S.M acknowledges funding from the Max Planck-Radboud University Center for Infrared Free-Electron Laser Spectroscopy. G.Á.-P. acknowledges support from the European Union (Marie Skłodowska-Curie Actions, grant agreement No. 101209198). N.S.M. acknowledges funding from the Deutsche Forschungsgemeinschaft (DFG, German Research Foundation) - Projektnummer 551280726. J.M.-S. acknowledges financial support from the Spanish Ministry of Science and Innovation (grant number PID2023148457NB-I00 funded by MCIN/AEI/10.13039/501100011033 and FSE +, CNS2024-154342 funded by MICIU/AEI/10.13039/501100011033. P.A.-G. acknowledges the Spanish Ministry of Science and Innovation under the National Plan grant PID2022-141304NB-I00, the European Research Council under Consolidator grant no. 101044461, TWISTOPTICS, and Agencia SEKUENS (Asturias) under grant UONANO IDE/2024/000678 with the support of FEDER funds.

\bibliographystyle{MSP}
\bibliography{bibliography}

\end{multicols}

\clearpage

\end{document}